\documentclass[fleqn,10pt]{wlscirep}
\usepackage[utf8]{inputenc}
\usepackage[T1]{fontenc}
\usepackage{lineno}
\usepackage{graphicx}
\usepackage{array}
\usepackage{longtable}
\usepackage{multirow}
\usepackage{tikz}
\usetikzlibrary{arrows,shapes,positioning}

\title{Accuracy Assessment of OpenAlex and Clarivate Scholar ID with an LLM-Assisted Benchmark}

\author[1,$\dag$]{Renyu Zhao}
\author[2,$\dag$]{Yunxin Chen}
\affil[1]{Tencent, SSV, Beijing, 100871, P.R.China}

\affil[*]{corresponding author(s): Renyu Zhao (roizhao@gmail.com)}

\affil[$\dag$]{these authors contributed equally to this work}

\begin{abstract}
    In quantitative SciSci (science of science) studies, accurately identifying individual scholars is paramount for scientific data analysis. However, the variability in how names are represented—due to commonality, abbreviations, and different spelling conventions—complicates this task. While identifier systems like ORCID are being developed, many scholars remain unregistered, and numerous publications are not included. Scholarly databases such as Clarivate and OpenAlex have introduced their own ID systems as preliminary name disambiguation solutions. This study evaluates the effectiveness of these systems across different groups to determine their suitability for various application scenarios. We sampled authors from the top quartile (Q1) of Web of Science (WOS) journals based on country, discipline, and number of corresponding author papers. For each group, we selected 100 scholars and meticulously annotated all their papers using a Search-enhanced Large Language Model method. Using these annotations, we identified the corresponding IDs in OpenAlex and Clarivate, extracted all associated papers, filtered for Q1 WOS journals, and calculated precision and recall by comparing against the annotated dataset.

\end{abstract}
\begin{document}

\flushbottom
\maketitle

\thispagestyle{empty}

\section{Introduction}
Accurate identification of individual researchers is a fundamental requirement in quantitative scientific research. The process of name disambiguation is essential for activities ranging from bibliometric analysis to scholarly networking. For example, in the reseach of international migration of scholars, each scholar should be labled carefully, especially because these scholars are migrations, their affiliation and other imformation might have changed overtime\cite{akbaritabar2024bilateral}. However, this task faces significant challenges due to the ambiguous nature of personal names. Researchers often share common names, and variations in name representation—such as initials, abbreviations, and different spelling conventions—can result in multiple representations of the same individual. Although unique identifier systems like ORCID have been proposed to address these issues, adoption remains incomplete, with many scholars yet to register and numerous journals not fully integrated into the system.

Commercial scholar databases such as Clarivate Analytics (with its Scholar ID system), Elsevier (Scopus), and open scholar databases such as OpenAlex have developed proprietary identification systems intended to resolve issues related to name ambiguity. These systems aim to unify the representation of individual scholars across various publications and databases by assigning unique identifiers. Among which Elsevier's Scopus have been roughly surveyed before\cite{10.1162/qss_a_00019}. However, not only it is not thorough for SciSci researchers to decide when their topics across different disciplines, regions, and levels of research activity, but also lack comparision of the effectiveness of other widely used scholar systems in accurately distinguishing and consolidating scholarly records.

To assess the accuracy of Scholar ID systems, a precise dataset of scholar and their publication list is needed, similar to that in the process of name disambiguating algorithms developing. Author self-maintained ID systems like ORCID, google scholar author page are commonly used\cite{Chen_BigScholar17}, yet might be biased since authors who maintain their profile on such systems might be a biased group. There are dataset prepared with complex and heavy datamining work out there\cite{10050437}\cite{10.1145/3219819.3219859}, yet they lack author profiles, which prevent us from assess the accuracies in different dimensions. To tackle all these problems, we employ LLM agents\cite{zhao2024scholardisambiguationsearchenhancedllm} to prepare a comprehensive dataset.

In this study a diversified sample of authors from the top tier (Q1) quartile of Web of Science (WOS) journals between year 2019 and 2023 is prepared. We consider three dimensions: country, discipline, and the number of papers published by the corresponding author. From each dimension's subset, 100 scholars were randomly selected, and all their papers were annotated by the agents mentioned above to serve as a reference dataset. Following this, we identified the scholars' OpenAlex and Clarivate IDs, retrieved all linked publications, and filtered these records to include only Q1 WOS journal articles. Precision and recall metrics were computed by comparing the annotated dataset with the publications linked to each ID system.

This research aims to provide a comprehensive evaluation of the disambiguation performance of OpenAlex and Clarivate Scholar IDs, offering insights into their applicability and reliability across various contexts. The outcomes will inform stakeholders about the current state of these systems and guide decisions regarding their deployment in different academic and analytical applications.

\section{Methodology}

\subsection{Scope and Preliminary Screening}

This study focuses on corresponding authors who published papers in Q1 journals between 2019 and 2023 across three major disciplines: Information and Electronics, Mathematics and Physics, and Life Sciences. The initial screening process applied the following criteria:
\begin{enumerate}
    \item journal domains were limited to Information and Electronics, Mathematics and Physics, and Life Sciences.
    \item Each corresponding author must have both a valid email address and institutional affiliation.
\end{enumerate}

\subsection{Data Collection and Initial Filtering}

Analysis of the initial dataset revealed that among 172 countries represented, the top ten countries account for 80.45\% of all corresponding authors. As illustrated in Table \ref{tab:science_fields}, China, the United States, and Europe emerge as primary regions for comparative analysis, given their substantial representation across all three disciplines.

\begin{table}[h]
    \centering
    \begin{tabular}{|c|c|c|c|c|c|}
    \hline
    \multicolumn{2}{|c|}{Life and Medical Sciences} & \multicolumn{2}{|c|}{Mathematics and Physics} & \multicolumn{2}{|c|}{Information and Electronics} \\ \hline
    \textbf{usa} & 54121 & \textbf{china} & 29797 & \textbf{china} & 27470 \\ \hline
    \textbf{china} & 44883 & \textbf{usa} & 11120 & \textbf{usa} & 14576 \\ \hline
    \textbf{england} & 10725 & \textbf{germany} & 3953 & \textbf{india} & 3100 \\ \hline
    \textbf{germany} & 8953 & \textbf{france} & 2910 & \textbf{england} & 2966 \\ \hline
    \textbf{france} & 6539 & \textbf{england} & 2755 & \textbf{germany} & 2693 \\ \hline
    \end{tabular}
    \caption{Publication counts by corresponding author country across three fields}
    \label{tab:science_fields}
\end{table}

To assess varying levels of scholarly activity, we categorized authors into three groups based on publication output: those with fewer than 3 papers, 3-5 papers, and more than 5 papers. This stratification was implemented following our disambiguation protocol.

\subsection{Author Disambiguation Process}

The disambiguation process began with email-based filtering, eliminating cases where a single name was associated with more than three email addresses. This initial step identified 207,107 email addresses, with 161,371 having unique name associations requiring no further disambiguation.

For cases where names were linked to multiple email addresses (two or three), we merged instances where the authors belonged to the same institution. This left 31,906 email addresses requiring additional disambiguation, while 167,417 maintained unique individual associations. Notably, Chinese authors comprised 69.18\% (22,074/31,906) of the multiple-email cases, while American authors accounted for 35.13\% (11,210/31,906).

We sampled 200 unique names from each module, with their associated emails forming our dataset. The country-specific datasets contained between 1,200 and 1,300 names, indicating that most author name were linked to two email addresses. The disambiguation process utilized cross-linguistic LLM agents \cite{zhao2024scholardisambiguationsearchenhancedllm}.

\subsubsection{Disambiguation Accuracy and Dataset Integrity}

While the search-enhanced LLM agents proved highly effective, we implemented strict quality control measures. Emails yielding no search-enhanced information were excluded from the standard dataset. The disambiguation agent evaluated whether authors sharing names but different emails were indeed the same individual. When confirmed, we consolidated their emails and research outputs. In cases where search-enhanced results were incomplete and could lead to ambiguous agent decisions, we retained only complete and consistent data with definitive 'False' disambiguation results.

Post-disambiguation analysis yielded 523 Chinese authors, 241 American authors, and 269 European authors. We capped each regional activity level module at 100 authors, supplementing with uniquely email-matched authors when necessary. The final standard datasets comprised 900 authors each for China, the USA and Europe.

\begin{tikzpicture}[
    node distance = 0.8cm and 1.2cm,
    auto,
    scale = 0.8,
    every node/.style={transform shape},
    >=stealth,
    font=\small,
    startstop/.style={rectangle, rounded corners, minimum width=2cm, minimum height=0.8cm,text centered, draw=black, fill=red!30},
    process/.style={rectangle, minimum width=2cm, minimum height=0.8cm, text centered, draw=black, fill=orange!30},
    decision/.style={rectangle, minimum width=2cm, minimum height=0.8cm, text centered, draw=black, fill=yellow!30},
    io/.style={trapezium, trapezium left angle=70, trapezium right angle=110, minimum width=2cm, minimum height=0.8cm, text centered, draw=black, fill=blue!30}
]

\node (start) [startstop] {Corresponding Authors of All Papers in Zone - 1 from 2019 - 2023};
\node (dec1) [decision, below=of start] {Are the Papers Belonging to the Three Major Disciplines?};
\node (discard) [startstop, right=4cm of dec1] {Discard};
\node (dec2) [decision, below=of dec1] {Are the Corresponding Authors' Email and Address Empty?};
\node (dec3) [decision, below=of dec2] {Filter Authors Whose Affiliated Institutions are in the Three Major Regions};
\node (process1) [process, below=of dec3] {Process in Sub - modules};
\node (dec4) [decision, below=of process1] {Is the Number of Same - named Email Addresses Greater than 3?};
\node (dec5) [decision, below=of dec4] {Is the Email Address in One - to - One Correspondence with the Person?};
\node (dec6) [decision, below=of dec5] {Are the Authors' Affiliated Institutions the Same?};
\node (process2) [process, below=of dec6] {Use LLM for Search Enhancement};
\node (dec7) [decision, below=of process2] {Is the Search Enhancement Result Empty?};
\node (dec8) [decision, below=of dec7] {Is the Search Enhancement Result Confirmed by the Disambiguation Agent?};
\node (merge) [process, below=of dec8] {Merge All Corresponding Email and Paper Information};
\node (fill_gap) [process, left=3cm of merge] {Fill in the Insufficient Part after Disambiguation};
\node (no_disambiguate) [process, left=3cm of dec6] {No Disambiguation Processing Required};
\node (dataset) [startstop, right=1cm of merge] {Dataset Obtained after Disambiguation};
\node (all_dataset) [startstop, below=of fill_gap] {All Datasets Obtained after Disambiguation};
\node (standard_dataset) [startstop, below=2cm of merge] {Standard Dataset};

\coordinate[right=1cm of dec4] (c1);
\coordinate[left=1cm of dec6] (c2);
\coordinate[below left=1cm and 1cm of dec8] (c3);

\draw[->] (start) -- (dec1);
\draw[->] (dec1) -- node[left] {Yes} (dec2);
\draw[->] (dec1) -| node[above,near start] {No} (discard.west);
\draw[->] (dec2) -- node[left] {No} (dec3);
\draw[->] (dec2) -| node[above] {Yes} (discard);
\draw[->] (dec3) -- node[left] {Yes} (process1);
\draw[->] (dec3) -| node[above] {No} (discard);
\draw[->] (process1) -- (dec4);
\draw[->] (dec4) -- node[left] {No} (dec5);
\draw[->] (dec4) -| node[above] {Yes} (discard);
\draw[->] (dec5) -- node[left] {No} (dec6);
\draw[->] (dec5) -| node[above] {Yes} (no_disambiguate);
\draw[->] (dec6) -- node[left] {No} (process2);
\draw[->] (dec6) -| node[above] {Yes} (no_disambiguate.east);
\draw[->] (process2) -- (dec7);
\draw[->] (dec7) -- node[left] {No} (dec8);
\draw[->] (dec7) -| node[above] {Yes} (discard);
\draw[->] (dec8) -- node[left] {Yes} (merge);
\draw[->] (dec8) -| node[above] {No} (discard);
\draw[->] (merge) -- (standard_dataset);
\draw[->] (no_disambiguate) |- (fill_gap.north);
\draw[->] (fill_gap) -- (all_dataset);
\draw[->] (all_dataset) -| (standard_dataset);
\draw[->] (dataset) |- (standard_dataset);
\draw[->] (dec8) -| node[pos=0.25,above] {Confirmed as Different Scholars} (dataset);

\end{tikzpicture}

\subsection{Data Definition}

Let \( S \) represent the complete set of scholars within our temporal scope. We randomly selected 900 scholars from each region (China, United States, and Europe). Formally, let \( S_{C} \subseteq S \) denote the subset of scholars from country/region \( C \), where:

\[
|S_{C}| = 900
\]

For each scholar \( s \in S_{C} \), designated agents conducted comprehensive manual annotation of all publications, establishing our reference dataset through meticulous review and verification.

\subsection{Data Processing}

Following annotation, we identified corresponding Clarivate IDs \(r\_id\) for each scholar. Let \( ID_{CL}(s) \) represent these identifiers. We then retrieved and filtered all associated publications to include only Q1 WOS journal articles.

Formally, with \( P_{CL}(s) \) representing the complete publication set linked to \( ID_{CL}(s) \), we define the filtered set \( P_{CL,Q1}(s) \) as:

\[
P_{CL,Q1}(s) = \{p \in P_{CL}(s) \mid p \text{ is a Q1 WOS journal article} \}
\]

\subsection{Metrics Computation}

To evaluate system performance, we computed precision and recall metrics comparing our annotated dataset against ID-linked publications. Let \( P_{ref}(s) \) represent the annotated paper set for scholar \( s \).

For each scholar \( s \), precision (\( \text{Prec} \)) and recall (\( \text{Rec} \)) are defined as:

\[
\text{Prec}_{CL}(s) = \frac{|P_{ref}(s) \cap P_{CL,Q1}(s)|}{|P_{CL,Q1}(s)|}
\]
\[
\text{Rec}_{CL}(s) = \frac{|P_{ref}(s) \cap P_{CL,Q1}(s)|}{|P_{ref}(s)|}
\]

Where:
- \( |P_{ref}(s)| \) is the number of papers a scholar has in the reference dataset (denoted as \( a \))
- \( |P_{CL,Q1}(s)| \) is the number of papers retrieved using r\_id (denoted as \( b \))
- \( |P_{ref}(s) \cap P_{CL,Q1}(s)| \) is the intersection of the two sets (denoted as \( c \))

This rigorous methodological framework ensures comprehensive and representative sampling, enabling robust cross-dimensional analysis of publication records.

\section{Experiments}

\subsection{Recall and Precision of Clarivate ID}

In this section, we evaluate the Clarivate data set by examining the recall and precision for scholars from China, the United States, and Europe. Each scholar is assigned a unique identifier \( r\_id \) by Clarivate. We first retrieve all \( r\_id \) corresponding to the papers in the standard dataset. Then, using these \( r\_id \), we obtain all the papers associated with each \( r\_id \) and compare them with the standard dataset.

Our findings indicate that among Chinese scholars, 147 individuals have multiple \( r\_id \), indicating an under-merge of scholar papers in the Clarivate database. Additionally, 40 \( r\_id \) are assigned to multiple individuals, indicating an over-merge of scholar papers. For American scholars, 104 individuals have multiple \( r\_id \), while 5 \( r\_id \) are assigned to multiple individuals. Among European scholars, 104 individuals have multiple \( r\_id \), and 171 \( r\_id \) are assigned to multiple individuals.

\begin{table}[ht]
\centering
\caption{Distribution of scholars with different numbers of \( r\_id \)}
\begin{tabular}{cccc}
\hline
Number of \( r\_id \) & Number of Chinese Scholars & Number of American Scholars & Number of European Scholars \\ \hline
0 & 30 & 14 & 9 \\
1 & 723 & 782 & 779 \\
2 & 130 & 96 & 104 \\
3 & 15 & 6 & 7 \\
4 & 2 & 2 & 1 \\ \hline
\end{tabular}
\end{table}

To evaluate the recall and precision from the perspective of individual scholars, we define the following calculation rules:
\[
\text{Recall} = \frac{c}{a}, \quad \text{Precision} = \frac{c}{b}
\]
where \( a \) is the number of papers a scholar has in the standard dataset, \( b \) is the number of papers retrieved using \( r\_id \), and \( c \) is the intersection of the two sets of papers.

From the analysis, 30 Chinese scholars lack \( r\_id \), resulting in a recall of 0. For 192 Chinese scholars, the recall values range between 0 and 1 due to some papers missing \( r\_id \) and hence not being recalled.

\begin{table}[ht]
\centering
\caption{Distribution of papers missing \( r\_id \)}
\begin{tabular}{cccc}
\hline
Number of Papers Missing \( r\_id \) & Number of Chinese Scholars & Number of American Scholars & Number of European Scholars \\ \hline
0 & 670 & 709 & 709 \\
1 & 147 & 137 & 132 \\
2 & 45 & 28 & 34 \\
3 & 11 & 6 & 10 \\
4 & 6 & 6 & 3 \\
5 & 7 & 5 & 5 \\
6 & 4 & 3 & 2 \\
7 & 2 & 0 & 1 \\
8 & 1 & 2 & 2 \\
9 & 2 & 0 & 0 \\
10 & 0 & 1 & 0 \\
11 & 3 & 1 & 1 \\
12 & 1 & 0 & 0 \\
14 & 1 & 0 & 0 \\
17 & 0 & 1 & 0 \\
18 & 0 & 0 & 1 \\
42 & 0 & 1 & 0 \\ \hline
\end{tabular}
\end{table}

\begin{figure}[htbp]
    \centering
    \includegraphics[width=0.8\linewidth]{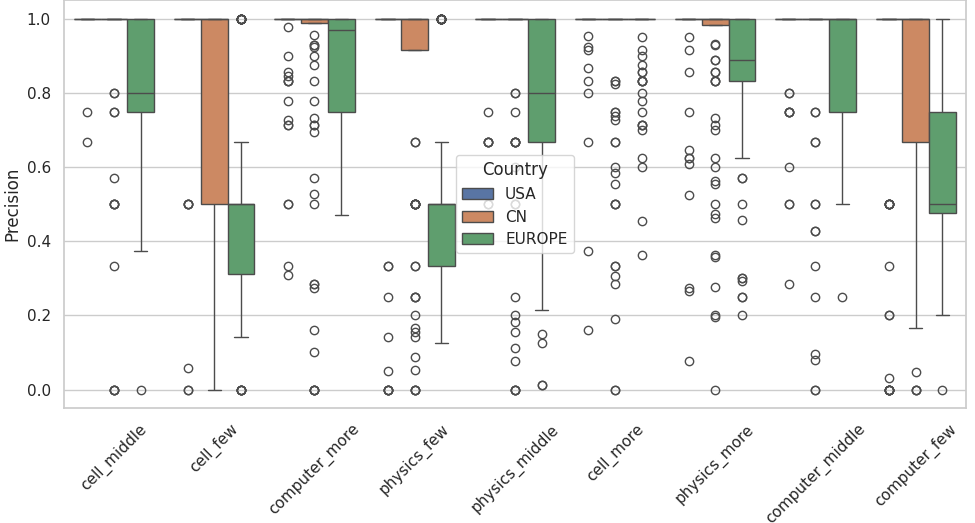}
    \caption{Precision of Clarivate ID.}
    \label{fig:boxplot-clarivate-precision}
\end{figure}

and the recall of Clarivate ID:
\begin{figure}[htbp]
    \centering
    \includegraphics[width=0.8\linewidth]{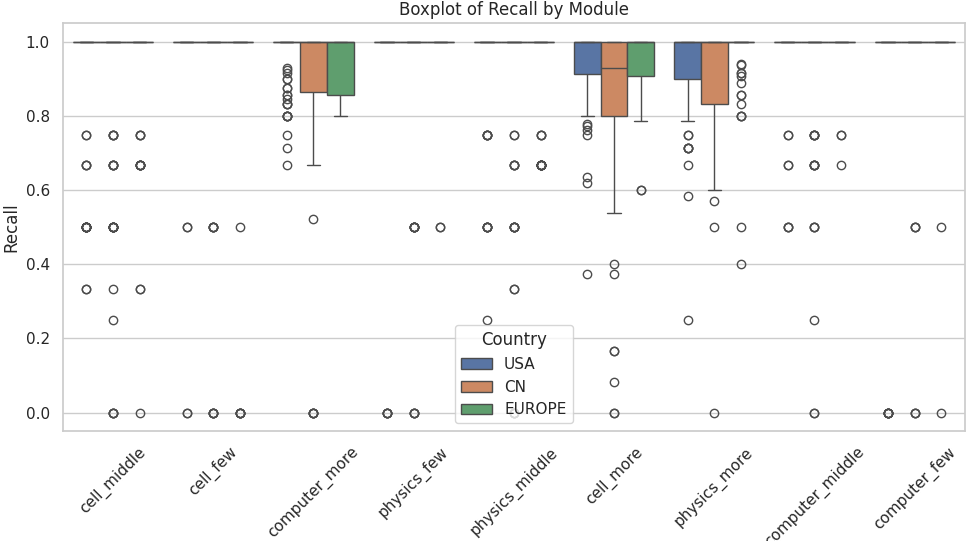}
    \caption{Recall of Clarivate ID.}
    \label{fig:boxplot-clarivate-recall}
\end{figure}

Figure \ref{fig:boxplot-clarivate-recall} shows the trends of recall across different regions. The recall rates for modules with scholars having five or more papers are dispersed. Conversely, recall rates for modules with fewer than five papers tend to cluster around 1.0, with only a few points below 1.0. Scholars with more papers are more likely to have some papers not appropriately assigned \( r\_id \), leading to more cases where recall < 1.

Chinese scholars often use pinyin formats, especially with two-character names, leading to frequent duplications. Consequently, modules with Chinese scholars having two or fewer papers show more variation in precision, reflecting a more dispersed distribution. Clarivate's handling of American scholars is notably better than other regions; modules across disciplines and paper count intervals do not exhibit large clusters below 1. Only a few errors are observed.

Overall, the Clarivate dataset matches American scholars and papers best, with both recall and precision significantly outperforming the other regions. Recall rates are generally higher than precision, indicating that instances of missing \( r\_id \) (under-merging) are less frequent than instances of incorrect assignment of \( r\_id \) (over-merging).

Next, we assess recall and precision at the module level using two measurement methods. The first method calculates the average recall and precision by averaging individual scholars' recall and precision within the module. The second method considers the entire module as a unit, calculating overall recall and precision. We denote the total number of papers in the standard dataset as \( a \), the number of papers retrieved via \( r\_id \) as \( b \), and the intersection as \( c \). Thus, the module-level recall and precision are defined as:
\[
\text{Module Recall} = \frac{c}{a}, \quad \text{Module Precision} = \frac{c}{b}
\]

The comparison in Figure \ref{fig:boxplot-clarivate-recall} shows that module recall and average recall produce nearly identical results, suggesting minimal impact of the calculation method on recall. However, there is a significant difference in precision between the two methods. Specifically, module precision curves follow similar trends with both showing the lowest and highest points at the same positions. However, the low points in module precision are lower than those in average precision.

\begin{figure}[htbp]
    \centering
    \includegraphics[width=0.8\linewidth]{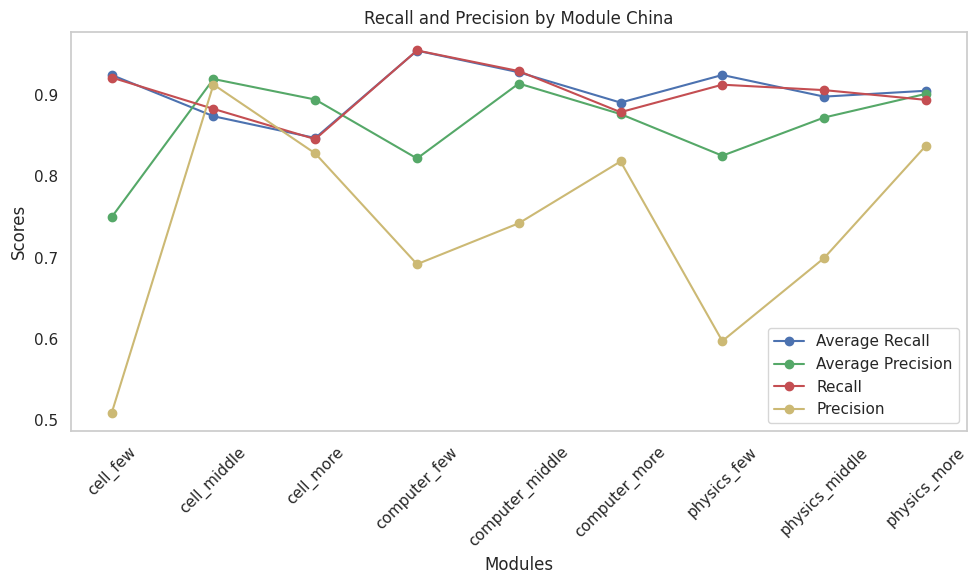}
    \caption{Precision and Recall of Clarivate ID in China.}
    \label{fig:china-clarivate}
\end{figure}

\begin{figure}[htbp]
    \centering
    \includegraphics[width=0.8\linewidth]{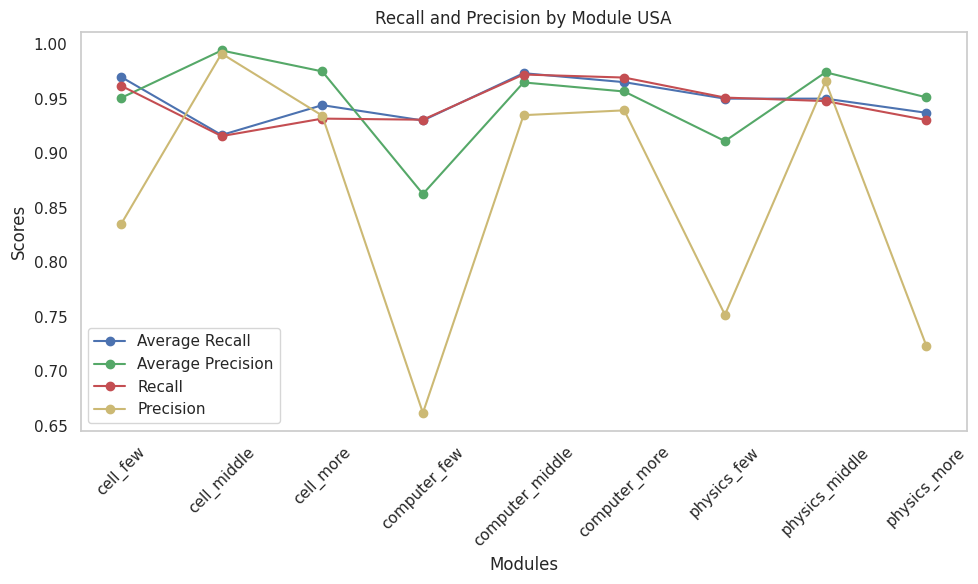}
    \caption{Precision and Recall of Clarivate ID in USA.}
    \label{fig:usa-clarivate}
\end{figure}

\begin{figure}[htbp]
    \centering
    \includegraphics[width=0.8\linewidth]{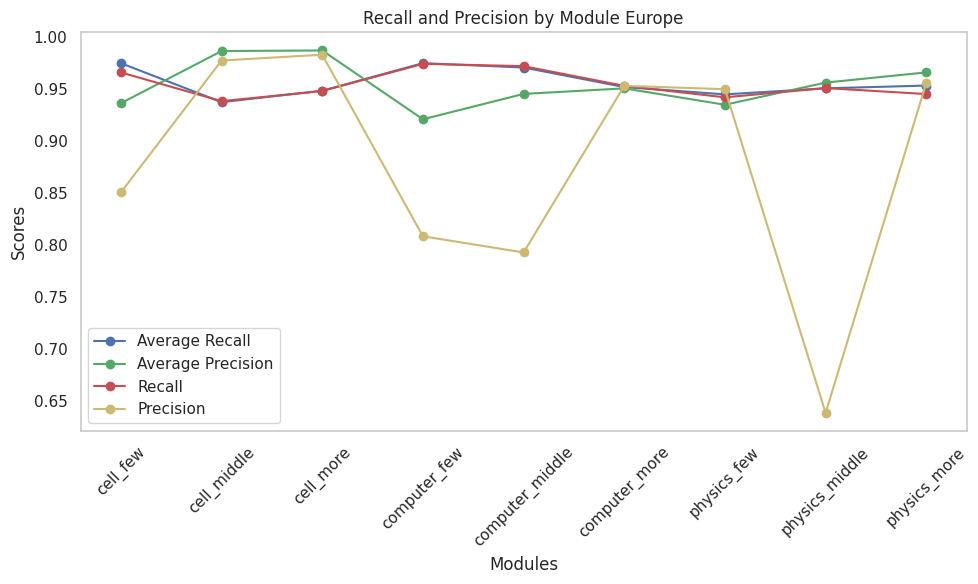}
    \caption{Precision and Recall of Clarivate ID in Europe.}
    \label{fig:europe-clarivate}
\end{figure}

\begin{figure}[htbp]
    \centering
    \includegraphics[width=0.8\linewidth]{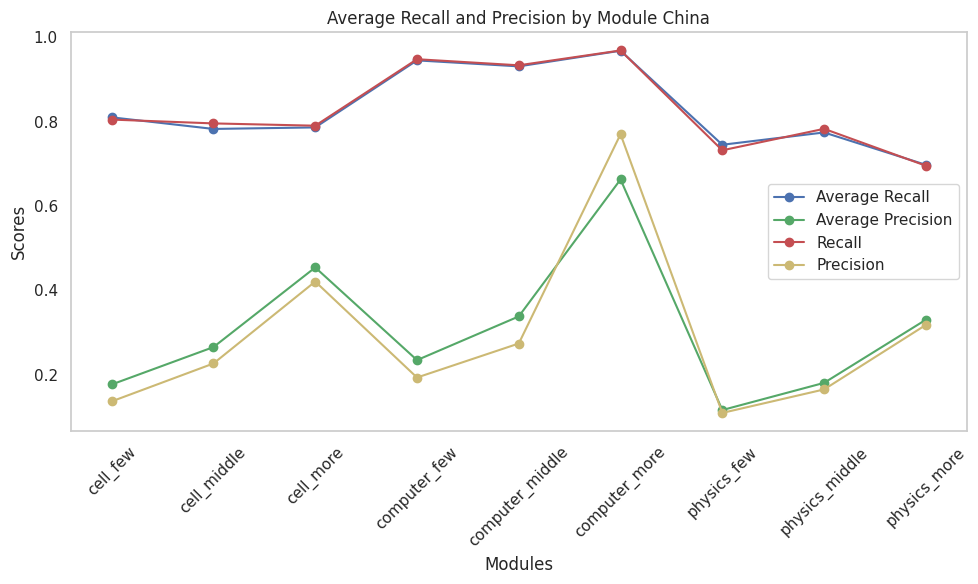}
    \caption{Precision and Recall of OpenAlex ID in China.}
    \label{fig:china-openalex}
\end{figure}

\begin{figure}[htbp]
    \centering
    \includegraphics[width=0.8\linewidth]{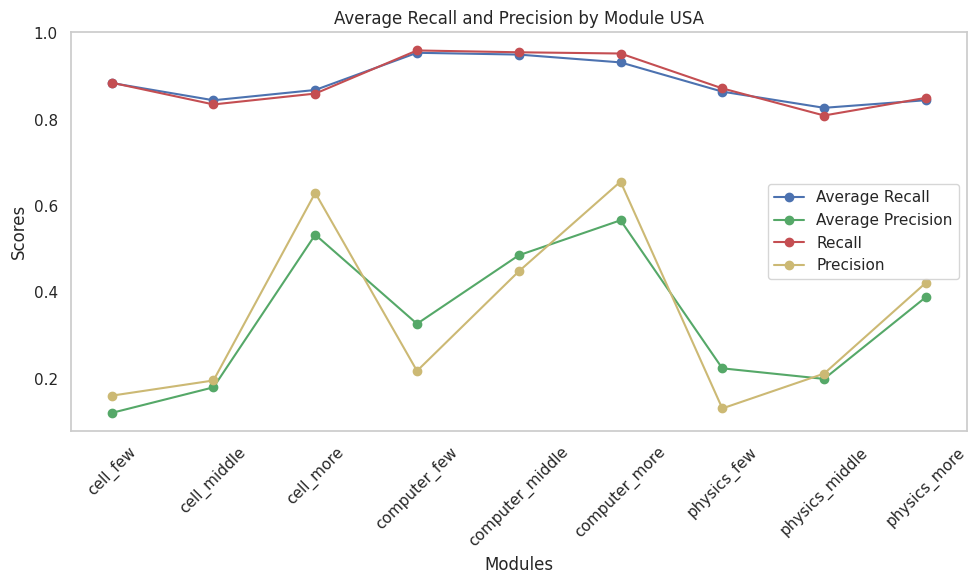}
    \caption{Precision and Recall of OpenAlex ID in USA.}
    \label{fig:usa-openalex}
\end{figure}

\begin{figure}[htbp]
    \centering
    \includegraphics[width=0.8\linewidth]{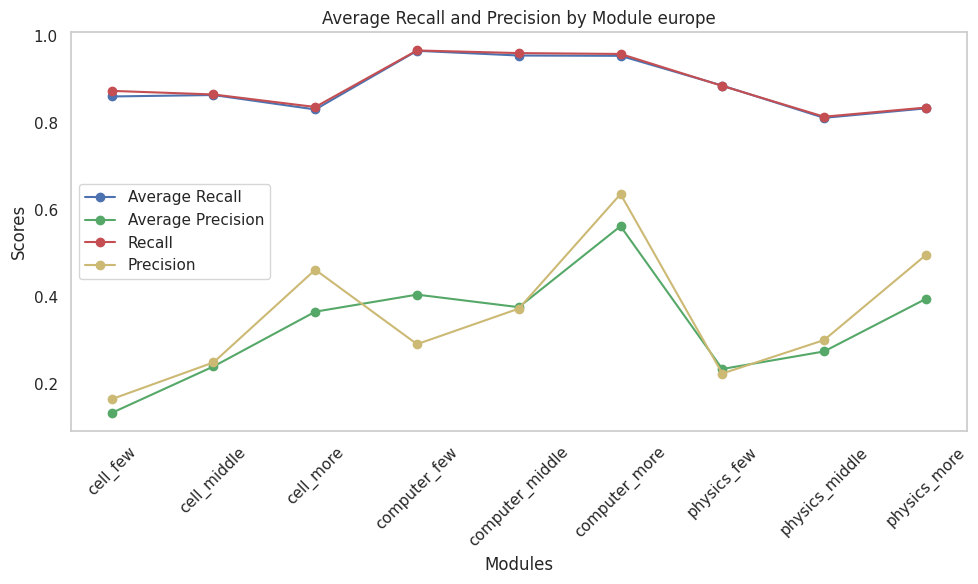}
    \caption{Precision and Recall of OpenAlex ID in Europe.}
    \label{fig:europe-openalex}
\end{figure}

\subsection{Recall and Precision of OpenAlex ID}

In this section, we evaluate the OpenAlex dataset using a standard dataset of 900 Chinese scholars whose work has been published in the first quartile over the past five years according to Clarivate Analytics. The evaluation process involves several steps as detailed below.

Firstly, we traverse all papers corresponding to each author to obtain their homepage addresses, also known as author\_id. If the author\_id cannot be retrieved through the papers, we search for the scholar's name along with their institutional information to acquire the corresponding author\_id. Using the author\_id, we then retrieve all their papers from OpenAlex, filtering only those published in the first quartile where the author is the corresponding author.

Due to inconsistent formatting of paper titles across databases, many exact matches might fail because of extraneous symbols. The average length of the paper titles was found to be 108 characters. In this experiment, we used an edit distance of 15; if the edit distance between two titles is less than or equal to 15, they are considered the same.

We employed two methods to find the intersection of papers. First, all papers obtainable through the OpenAlex API were included in the intersection, which resulted in 851 scholars having non-empty paper sets in OpenAlex. Second, the intersection was taken between the standard dataset and the papers retrieved via author\_id, resulting in 657 scholars with non-empty intersections. It should be noted that some papers in OpenAlex can be queried but do not appear on the author's homepage due to incorrect disambiguation or classification. Therefore, the first method is considered more representative.

Issues were identified with the OpenAlex dataset:
\begin{enumerate}
    \item Taking the intersection of papers obtained from the standard dataset and OpenAlex, 463 papers were analyzed for detailed information such as publication year, number of authors, author's position in the author list, and affiliated institutions. Discrepancies in publication year were found in 162 papers when compared to Clarivate Analytics, verified to be correct by Google Scholar and other publication databases, indicating significant errors in basic information within OpenAlex.
    \item An unexpected indexing error in OpenAlex was discovered: a paper (let's call it 'a') listed on an author's homepage could not be retrieved via the API.
    \item We also found inadequacies in the disambiguation process. Identical papers appeared under multiple paper\_ids. For instance, the paper titled "Glycosyltransferase UGT79B7 negatively regulates hypoxia response through $\gamma$ -aminobutyric acid homeostasis in Arabidopsis" had two results in OpenAlex: \url{https://openalex.org/W3135641188} and \url{https://openalex.org/W3194753228}.
\end{enumerate}

Among the 900 Chinese scholars, 851 had non-empty intersections. 87 scholars had multiple author\_id links, 72 lacked author\_id, and 176 experienced multiple returns for the same title. The most frequent case showed 8 out of 23 papers being duplicated. For 900 American scholars, 869 had non-empty intersections, 54 had multiple author\_ids, 108 lacked author\_ids, and 391 experienced multiple returns for the same title.

\begin{figure}[htbp]
    \centering
    \includegraphics[width=0.8\linewidth]{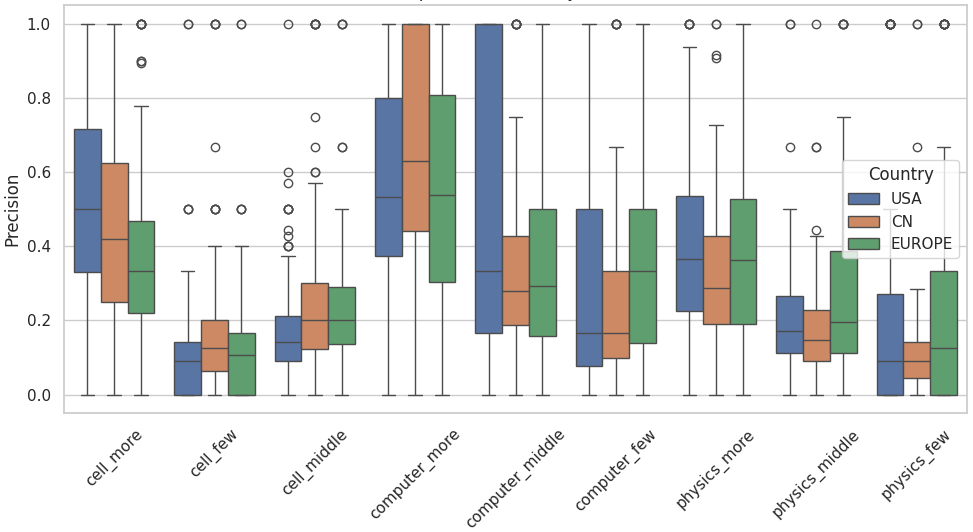}
    \caption{Precision of OpenAlex ID.}
    \label{fig:boxplot-openalex-precision}
\end{figure}

\begin{figure}[htbp]
    \centering
    \includegraphics[width=0.8\linewidth]{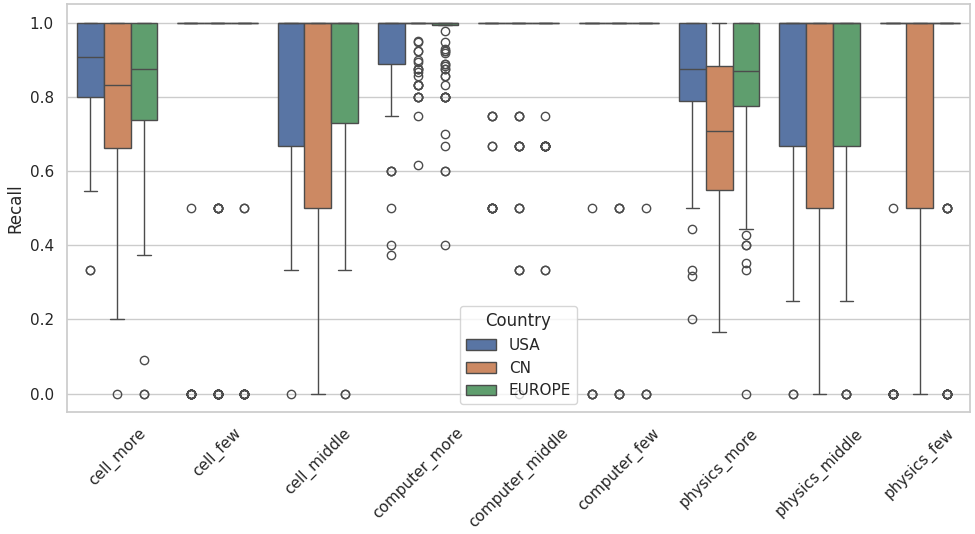}
    \caption{Recall of OpenAlex ID.}
    \label{fig:boxplot-openalex-recall}
\end{figure}

As shown in Figure \ref{fig:boxplot-openalex-precision} and Figure \ref{fig:boxplot-openalex-recall}, the precision across various regions is not high, indicating challenges faced by OpenAlex in accurately disambiguating scholars with identical names. When extracting all institutions for a single scholar, if a scholar has affiliations a and b for paper A, and a and c for paper B, we consider a, b, and c as belonging to the same author. For example, under "zhang, zhiguo", there are three unrelated institution sets, likely indicating three different individuals with the same name.

Additionally, a combined comparison among three regions reveals that the recall rate in the information electronics field exceeds 90\%, significantly higher than in the other two fields. This could be attributed to broader collaboration and more active research in the information electronics domain, leading to better classification outcomes.

\section*{Usage Notes}

The dataset provided in this study includes publicly available information sourced from web pages, specifically focusing on author details from other academic publications. This data is intended solely for academic and research purposes to support reproducibility and transparency of scientific research.

Researchers using this dataset are encouraged to respect the original context and purpose of the data. Any secondary use of the dataset should adhere to ethical guidelines and best practices in data handling, including proper citation of the original sources.

While the data included in this dataset is publicly accessible, it is important to acknowledge potential privacy considerations. The collection, use, and dissemination of this information have been conducted with the intent to respect privacy and intellectual property rights.

Users of this dataset must ensure compliance with relevant data protection regulations, institutional policies, and ethical standards. Neither the authors of this study nor their affiliated institutions assume responsibility for any misuse of the data or any legal implications that may arise from its use. It is the responsibility of the end-users to ensure that their usage of the dataset does not infringe upon the privacy or rights of individuals.

\section*{Code availability}
Data and code will be available at \href{https://github.com/sylviachen0513/}{github.com/sylviachen0513}.

\bibliography{ref}

\section*{Author contributions statement}
R.Z. conceived the experiment(s), Y.C. and R.Z. conducted the experiment(s). All authors reviewed the manuscript. 

\section*{Competing interests}

The corresponding author is responsible for providing a \href{https://www.nature.com/sdata/policies/editorial-and-publishing-policies#competing}{competing interests statement} on behalf of all authors of the paper.

\end{document}